\documentclass[12pt]{article}
\usepackage{graphicx}

\newcommand\pubnumber{Article 37 in
eConf C1304143}
\newcommand\pubdate{\today}

\def\Title#1{\begin{center} {\Large #1 } \end{center}}
\def\Author#1{\begin{center}{ \sc #1} \end{center}}
\def\Address#1{\begin{center}{ \it #1} \end{center}}

\newcommand\pubblock{\rightline{\begin{tabular}{l} \pubnumber\\
         \pubdate  \end{tabular}}}
\newenvironment{Abstract}{\begin{quotation}  }{\end{quotation}}
\newenvironment{Presented}{\begin{quotation} \begin{center}
             PRESENTED AT\end{center}\bigskip
      \begin{center}\begin{large}}{\end{large}\end{center} \end{quotation}}
\def\Acknowledgements{\bigskip  \bigskip \begin{center} \begin{large}
             \bf ACKNOWLEDGEMENTS \end{large}\end{center}}
\def\beq{\begin{equation}}
\def\eeq#1{\label{#1}\end{equation}}
\def\eeqn{\end{equation}}

\def\beqa{\begin{eqnarray}}
\def\eeqa#1{\label{#1}\end{eqnarray}}
\def\eeqan{\end{eqnarray}}

\let\bar=\overbar

\def\Dslash{\not{\hbox{\kern-4pt $D$}}}
\def\dslash{\not{\hbox{\kern-2pt $\del$}}}

\def\msb{{\bar{\ssstyle M \kern -1pt S}}}

\begin{document}
\begin{titlepage}
\pubblock

\vfill
\Title{The highest redshift gamma-ray bursts}
\vfill
%\Author{ Nial Tanvir\support}
\Author{ Nial Tanvir}
\Address{Department of Physics and Astronomy, University of Leicester, Leicester, LE1 7RH, United Kingdom}
\vfill
\begin{Abstract}
I review the searches for gamma-ray bursts at very high redshift.
Although the numbers of GRBs known at $z>6$ remain few,
even small samples can provide information about early star and galaxy formation
in the universe which is very hard to
obtain by any other means.

\end{Abstract}
\vfill
\begin{Presented}
GRB 2013 \\
the Seventh Huntsville Gamma-Ray Burst Symposium \\
Nashville, Tennessee, 14--18 April 2013
\end{Presented}
\vfill
\end{titlepage}
\def\thefootnote{\fnsymbol{footnote}}
\setcounter{footnote}{0}

\section{Introduction}

The formation of early collapsed structures in the universe, and the processes
of enrichment and reionization that they initiated, are topics of intense current interest.
Detection of individual galaxies at $z>7$ has proven highly challenging and, although
moderately large samples of candidates have been identified using a combination
of {\em Hubble Space Telescope}  and {\em Spitzer Space Telescope} photometry \cite{ellis13},
it still appears likely that they include low-$z$ interlopers \cite{pirzkal13}.
In addition, the evidence suggests that the dominant proportion of star formation
at these times was taking place in galaxies below the {\em HST} detection limit \cite{robertson13},
limiting the  conclusions that can be drawn.

It has long been argued that long-duration gamma-ray bursts (GRBs), thanks
to their extreme luminosities and association with massive star death (and hence
star formation) are potentially powerful probes of the early universe, providing
access to information that is hard to obtain in other ways (e.g., \cite{lamb00,tanvir07}).

\section{Discovery of high redshift GRBs}

High redshift ($z>6$) GRBs have generally been identified first as candidates
via optical and infrared (nIR) photometry of their afterglows, where they stand out as dropout sources 
in bluer bands, ideally together with a blue continuum slope in the redder bands \cite{haislip06}.
Spectroscopic detection of the Lyman-$\alpha$ break provides a clear feature from which
definitive redshifts are obtained.  In this way, several GRBs have now been confirmed
at $z>6$ \cite{kawai06,greiner09,tanvir09,salvaterra09}.

There have also been instances where no spectrum was obtained, due to poor
weather conditions and/or limited availability of nIR spectrographs.  In these
cases, photometric redshifts can provide strong constraints, albeit with
larger uncertainties.  The most notable example is that of GRB\,090429B
with a photo-$z\approx9.4$, although with some dust contribution to the reddening, this could have been as low as $z\sim7$
\cite{cucchiara11}.
In this regard, GRB afterglows have the advantage over galaxies of having simple
power-law spectra, and so photometric redshifts are generally more robust and not
subject to catastrophic errors \cite{kruehler11}.

\begin{figure}[htb]
\centering
\includegraphics[height=7cm]{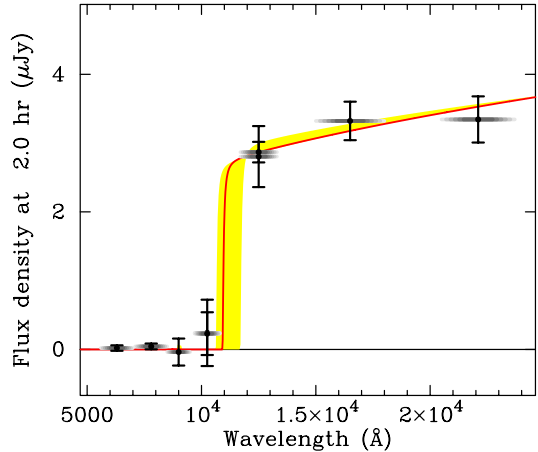}
\caption{Spectral energy distribution of the afterglow of GRB\,120923A, showing it
to be best fit (red line) by a power-law cut off by a Lyman-$\alpha$ break at $z\approx8.5$.
Note, photometric data have been extrapolated to a common time using the
measured broken power-law light curve.}
\label{923A}
\end{figure}

A recent very high redshift burst is GRB\,120923A.  Work on the spectra
of that event is ongoing, but a provisional analysis of the photometric data
indicates $z\approx8.5$, as shown in Figure~\ref{923A}.

%%%%%%%%%%%%%%%%%%%%%%%%%%%%%%%%%%%%%%%%%%%%%%%%%%%%%%%%%%%%%%%%%%%%%%%%%%%

\section{The potential of spectroscopy}

Afterglow spectra can also provide key diagnostics of the environments of the
bursts.  
Only the red wing of the Lyman-$\alpha$ line is typically seen at high redshift,
due to the continuum to the blue being absorbed by neutral hydrogen in the intergalactic
medium (IGM).  However, with a sufficiently high signal-to-noise spectrum, and a redshift
precisely determined via metal absorption lines, the red wing can be fitted simultaneously by
an IGM component along with a component due to the interstellar medium in the host galaxy.
Both of these components potentially provide key diagnostics.
The former gives  the IGM neutral fraction proximate to the host, and thus
a measure of the progress of reionization; if this can be achieved for a sample of
GRBs at different redshifts, then the time history and sight-line to sight-line variance of reionization can be constrained \cite{mcquinn08}.
The latter, the HI column due to the host, can be used to infer the opacity of the gas to
ionising radiation, and hence the escape fraction.  Once again, if this can be achieved for
a substantial sample of GRBs, then the distribution of escape fractions to high-$z$
star-formation would be deduced \cite{fynbo09}.  This is important since a high escape fraction is required
in order for stars to provide the necessary photon budget to drive reionization.
We note that the host HI absorption is also required in order to translate the column
densities of metal lines into abundances relative to hydrogen, but once achieved, this can provide
unique insights into chemical enrichment in star forming galaxies at $z>2$, e.g. \cite{thoene13}.

So far there has been rather limited success in pursuing this goal at $z>6$, due to the low
rate of high redshift GRBs in general, and of intrinsically bright afterglows in particular.
However, an early success with GRB\,050904 at $z=6.3$, pointed to the potential
of this approach \cite{totani06}, and GRB\,130606A looks likely to provide a further
opportunity to apply these techniques.  In the long run, spectroscopy of even faint high-$z$ GRB afterglows 
with 30\,m class telescopes should enable exquisitely precise measurements to be 
obtained.

\section{The population of high redshift GRBs}

\begin{figure}[htb]
\centering
\includegraphics[height=13.5cm]{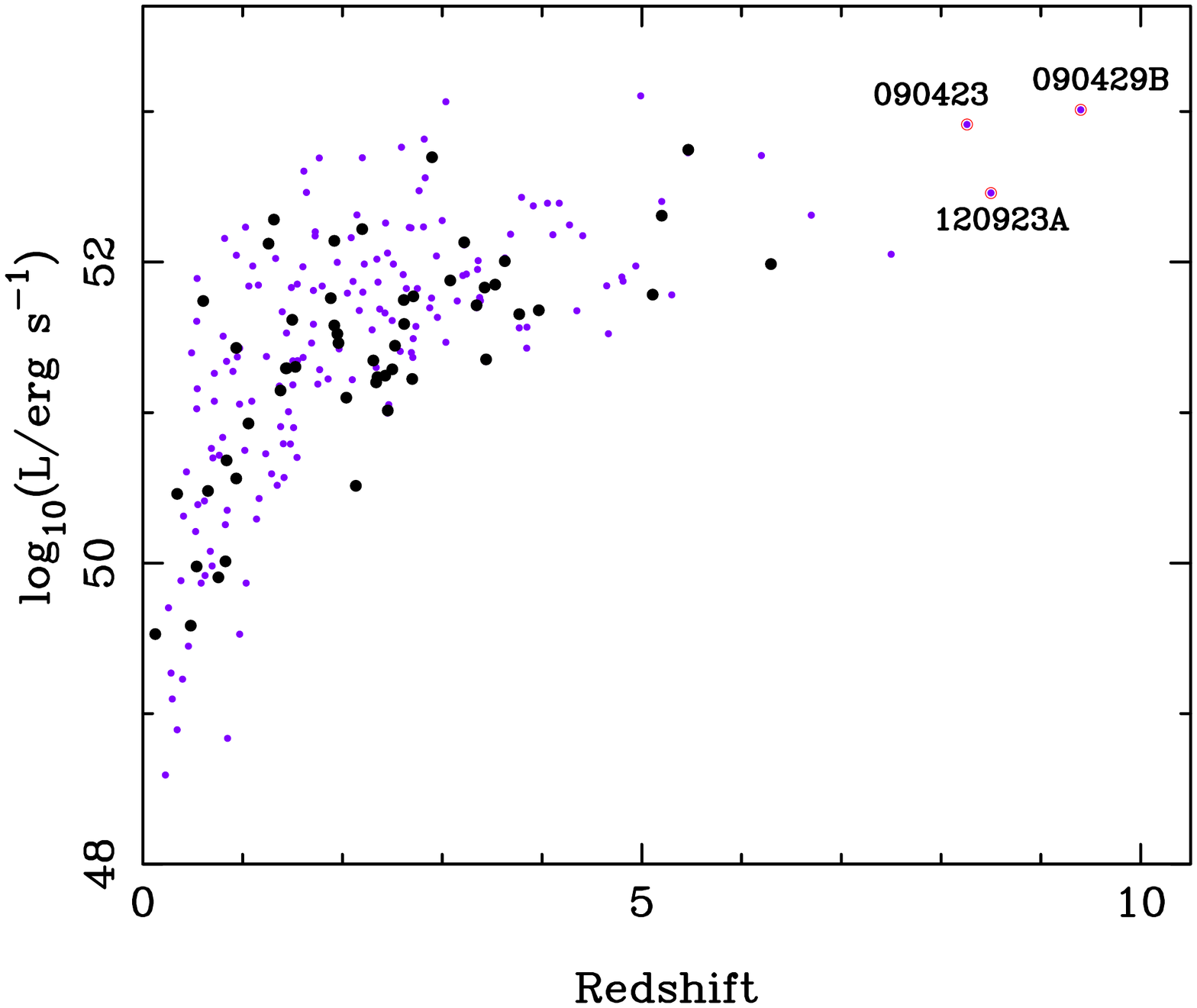}
\caption{Plot of peak luminosity versus redshift, for all {\em Swift} bursts
where a redshift has been measured.  Several of the highest redshift bursts
have been labelled, and those in the TOUGH sample \cite{hjorth12} (which has a high
degree of redshift completeness, and therefore small optical bias) shown as bold symbols.
The lower envelope is imposed by the BAT detection threshold, although we note that
in practice there are a large number of trigger algorithms used, and so the selection 
limit is not simply a function of peak luminosity.}
\label{l_vs_z}
\end{figure}

It is interesting to ask how the GRBs found at high redshift compare to the lower redshift
population.  Of course, bright bursts with bright afterglows are more likely to be localised
by {\em Swift} and have their redshifts measured via optical/nIR spectroscopy.
The selection of bright bursts is illustrated in Figure~\ref{l_vs_z}, which shows the intrinsic
peak luminosity ($k$-corrected to a common energy band of 30--300\,keV using the
observed spectrum) versus redshift for the sample of {\em Swift} GRBs with redshifts. 

%%%%%%%%%%%%%%%%%%%%%%%%%%%%%%%%%%%%%%%%%%%%%%%%%%%%%%%%%%%%%%%%%%%%%%%%%%%

\begin{figure}[htb]
\centering
\includegraphics[height=18cm]{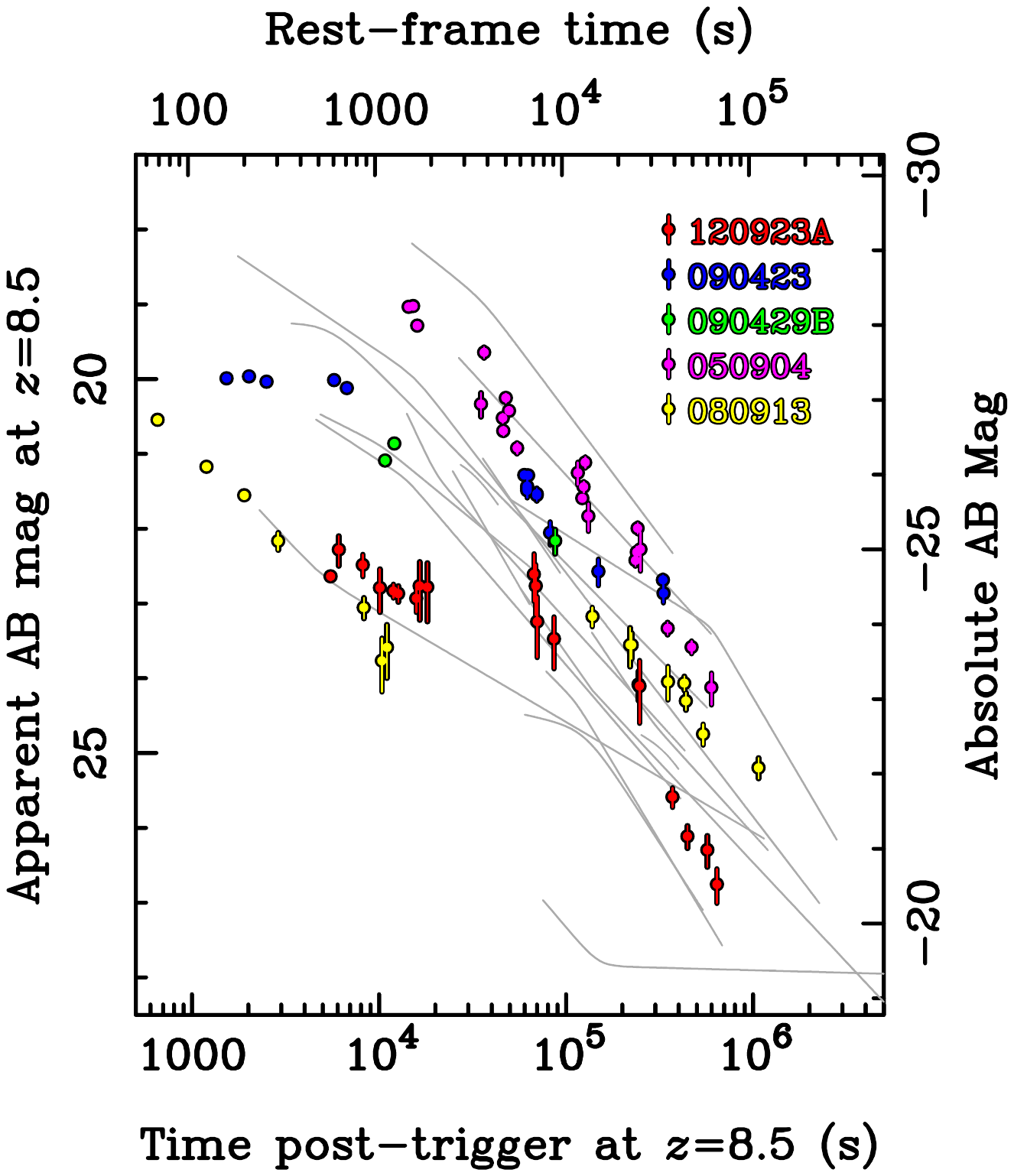}
\caption{The rest-frame optical light curves of a large number of GRBs,
artificially redshifts to show how they would appear at $z=8.5$ (grey lines
are broken power-law model fits to data from \cite{kann10}), together with
individual data points \cite{haislip06,greiner09,tanvir09,cucchiara11} for several of the known highest redshift GRBs (also
shifted to $z=8.5$).}
\label{lco}
\end{figure}

In order to precisely localise and obtain a redshift it is also required that the bursts
have intrinsically bright afterglows in the observed nIR.
Figure~\ref{lco} shows that the five highest redshift bursts in fact span a fairly
large range in brightness compared to a sample of optically observed
afterglows at lower redshift (although this comparison sample itself is biased
against including particularly faint afterglows).
This suggests that the selection effect due to limitations of ground observations
is not the overriding consideration, although doubtless some bursts detected
by {\em Swift} are missed due to insufficient opportunity for ground follow-up.

%%%%%%%%%%%%%%%%%%%%%%%%%%%%%%%%%%%%%%%%%%%%%%%%%%%%%%%%%%%%%%%%%%%%%%%%%%%

Generally, the high-$z$ GRB sample seems to be drawn from the same
population as observed at lower-$z$, although see Littlejohns et al. (this
conference) for indications that the rest-frame durations of $z>5$ GRBs
may be on average shorter.

\section{The host population}

To date, there have been no detections of $z>6$ GRB hosts in deep imaging,
which is consistent with the bulk of star formation in the reionization era occurring 
in faint galaxies, below the effective detection limit of {\em HST} \cite{tanvir12}.
It will clearly be of considerable interest to target the sample of $z>6$ GRB hosts
with {\em JWST} when it is launched.

\section{Conclusions}

The sample of $z>6$ GRBs continues to grow, and are beginning to provide
interesting results impacting on early star formation and galaxy evolution.
Overall, numbers remain small, though, and while difficulties of followup likely mean that some proportion of {\em Swift}
detected high-$z$ events are being missed, the rest-frame properties of the observed
sample suggest that even faint and difficult afterglows are being found.
All this motivates new missions dedicated to locating high-$z$ GRBs.

%Recently, F.  A. Mesmer \index{Mesmer}
%physiology~\cite{Mesmer}.
%Numerous authors have claimed to reproduce certain of the phenomena
%reported by Mesmer~\cite{diCenzo}, but there.

%My observations began when I was called on an sensibilities 
%\index{sensibilities}  of two

%The in Figure~\ref{fig:magnet}.  The device Table~\ref{tab:blood}.

%%%%%%%%%%%%%%%%%%%%%%%%%%%%%%%%%%%%%%%%%%%%%%%%%%%%%%%%%%%%%%%%%%%%%%%%%
%%
%%   use this format to include a LaTeX table  into your paper
%%
%\begin{table}[t]
%\begin{center}
%\begin{tabular}{l|ccc}
%Patient &  Initial level($\mu$g/cc) &  w. Magnet &
%w. Magnet and Sound \\ \hline
% Guglielmo B.  &   0.12     &     0.10      &     0.001  \\
% Ferrando di N. &  0.15     &     0.11      &  $< 0.0005$ \\ \hline
%\end{tabular}
%\caption{Blood cyanide levels for the two patients.}
%\label{tab:blood}
%\end{center}
%\end{table}
%%%%%%%%%%%%%%%%%%%%%%%%%%%%%%%%%%%%%%%%%%%%%%%%%%%%%%%%%%%%%%%%%%%%%%%%%%%

%\section{Interpretation}

%A description  L. da Ponte~\cite{daPonte}. \index{da Ponte} In his paper,

\Acknowledgements
I am grateful to my many collaborators in high-$z$ GRB related projects; and in relation to
the projects highlighted here, particularly
Andrew Levan, Klaas Wiersema, Dan Perley, Jens Hjorth, Johan Fynbo, Daniele Malesani,
Palli Jakobsson, Thomas Kr\"uhler, Brad Cenko, Nino Cucchiara, Johannes Zabl, Alex Kann, Tanmoy Laskar.


\begin{thebibliography}{99}

%%
%%  bibliographic items can be constructed using the LaTeX format in SPIRES:
%%    see    http://www.slac.stanford.edu/spires/hep/latex.html
%%  SPIRES will also supply the CITATION line information; please include it.
%%


\bibitem{ellis13} Ellis R.~S., et al., ApJ, 763, L7 (2013).
\bibitem{pirzkal13} Pirzkal N., Rothberg B., Ryan R., Coe D., Malhotra S., Rhoads J., Noeske K., arXiv:1304.4594 
\bibitem{robertson13} Robertson B.~E., et al., ApJ, 768, 71 (2013).
\bibitem{lamb00} Lamb D.~Q., Reichart D.~E., ApJ, {\bf 536}, 1 (2000).
\bibitem{tanvir07} Tanvir N.~R., Jakobsson P.,  RSPTA, {\bf 365}, 1377 (2007).
\bibitem{haislip06} Haislip J.~B., et al., Nature, {\bf 440}, 181 (2006).
\bibitem{kawai06} Kawai N., et al., Nature, {\bf 440}, 184 (2006).
\bibitem{greiner09} Greiner J., et al., ApJ, {\bf 693}, 1610 (2009).
\bibitem{tanvir09} Tanvir N.~R., et al.,  Nature, {\bf 461}, 1254 (2009).
\bibitem{salvaterra09} Salvaterra R., et al., Nature, {\bf 461}, 1258 (2009).
\bibitem{cucchiara11} Cucchiara A., et al., ApJ, {\bf 736}, 7 (2011).
\bibitem{kruehler11} Kr{\"u}hler T., et al., A\&A, 526, A153 (2011).
\bibitem{mcquinn08} McQuinn M., et al., MNRAS, {\bf 388}, 1101 (2008).
\bibitem{fynbo09} Fynbo J.~P.~U., et al., ApJS, {\bf 185}, 526 (2009).
\bibitem{thoene13} Th{\"o}ne C.~C., et al., MNRAS, {\bf 428}, 3590 (2013).
\bibitem{totani06} Totani T., et al., PASJ, {\bf 58}, 485 (2006).
\bibitem{hjorth12}  Hjorth J., et al., ApJ, {\bf 756}, 187 (2012).
\bibitem{tanvir12} Tanvir N.~R., et al., ApJ, {\bf 754}, 46 (2012).
\bibitem{kann10} Kann D.~A., et al., ApJ, {\bf 720}, 1513 (2010). 

%\bibitem{conselice05} Conselice C.~J., et al., ApJ, {\bf 633}, 29 (2005).
%\bibitem{fynbo06} Fynbo J.~P.~U., et al., A\&A, {\bf 451}, L47 (2006).
%\bibitem{greiner11} Greiner J., et al., A\&A, {\bf 526}, A30 (2011).
%\bibitem{perley09} Perley D.~A., et al., AJ, {\bf 138}, 1690 (2009).
%\bibitem{ruiz07} Ruiz-Velasco A.~E., et al., ApJ, {\bf 669}, 1 (2007).
%\bibitem{tanvir08} Tanvir N.~R., et al., MNRAS, {\bf 388}, 1743 (2008).
%%\bibitem{zafar11} Zafar T., Watson D.~J., Tanvir N.~R., Fynbo J.~P.~U., Starling R.~L.~C., Levan A.~J., ApJ, {\bf 735}, 2 (2011).

\end{thebibliography}
\end{document}